\documentclass[a4paper,10pt]{article}

\usepackage[ansinew]{inputenc}
\usepackage{amssymb, amsthm, bbm}
\usepackage{latexsym}
\usepackage{amsmath}
\usepackage{hyperref}
\usepackage[dvips]{graphicx}

\newcommand{\und}{\qquad\text{and}\qquad}
\newcommand{\sfrac}[2]{{\textstyle\frac{#1}{#2}}}
\renewcommand{\=}{\ =\ }
\newcommand{\tr}{\text{tr}}
\newcommand{\unity}{\mathbbm{1}}

\theoremstyle{definition}

\makeatletter
\renewcommand{\@maketitle}{
\newpage
\null
\begin{flushright}
ITP--UH--04/12\\
\end{flushright}
\vskip 1em%
\begin{center}%
{\Large \textbf \@title \par}%
\vskip 2.5em 
{\large \@author \par}
\vskip 1.5 em
\end{center}%
\par} \makeatother

\title{
Nearly K\"{a}hler heterotic compactifications \\ with fermion condensates
}
\author{
Athanasios Chatzistavrakidis$^{\ \$}$, \
Olaf Lechtenfeld$^{\ \dagger\times}$ and \ 
Alexander D. Popov$^{\ \ast}$
}
\date{}

\begin{document}
\setcounter{page}{0}
\parindent=0cm
\maketitle
\thispagestyle{empty}

\begin{center} 

${}^{\$}${\em 
Bethe Center for Theoretical Physics and Physikalisches Institut der Universit\"{a}t Bonn,\\
Nussallee 12, D-53115 Bonn, Germany } \\ e-mail: than@th.physik.uni-bonn.de

\bigskip

${}^\dagger${\em 
Institut f\"ur Theoretische Physik, Leibniz Universit\"at Hannover \\
Appelstra{\ss}e 2, 30167 Hannover, Germany } \\
e-mail: lechtenf@itp.uni-hannover.de

\bigskip

${}^\times${\em
Centre for Quantum Engineering and Space-Time Research \\
Leibniz Universit\"at Hannover \\
Welfengarten 1, 30167 Hannover, Germany }

\bigskip

$^\ast${\em Bogoliubov Laboratory of Theoretical Physics, JINR \\
141980 Dubna, Moscow Region, Russia} \\
e-mail: popov@theor.jinr.ru
\end{center}

\bigskip

\begin{abstract} 
We revisit AdS${}_4$ heterotic compactifications on nearly K\"ahler manifolds 
in the presence of $H$-flux and certain fermion condensates. Unlike previous
studies, we do not assume the vanishing of the supersymmetry variations. 
Instead we determine the full equations of motion originating from the 
ten-dimensional action, and subsequently we provide explicit solutions to them
on nearly K\"ahler manifolds at first order in $\alpha'$. The Bianchi identity
is also taken into account in order to guarantee the absence of all anomalies.
In the presence of $H$-flux, which is identified with the torsion of the 
internal space, as well as of fermion condensates in the gaugino and dilatino
sectors, new solutions are determined. These solutions provide a full 
classification of consistent backgrounds of heterotic supergravity under our 
assumptions. All the new solutions are non-supersymmetric, while previously
known supersymmetric ones are recovered too. Our results indicate that fully 
consistent (supersymmetric or not) heterotic vacua on nearly K\"ahler manifolds
are scarce, even on AdS${}_4$, and they can be completely classified.
\end{abstract}

\newpage

\numberwithin{equation}{section}

\section{Introduction}

Calabi-Yau compactifications of string theory, despite their welcome features, suffer from the infamous moduli 
problem, the existence of flat directions in the four-dimensional potential corresponding to scalar fields 
which are not stabilized. The most plausible scenario which resulted from the attempts to resolve this problem 
was the introduction of fluxes, namely vacuum expectation values of the tensor fields of the theory along 
the compactification manifold. These fluxes typically lead to a deformation of the internal manifold 
away from the Calabi-Yau property and suggest the study of non-K\"ahler manifolds 
\cite{Strominger:1986uh,LopesCardoso:2002hd}.

Unlike type II string theories, where the plentitude of Ramond-Ramond fields offers considerable freedom in 
the introduction of internal fluxes, the heterotic string case is more restrictive. Indeed, in the heterotic 
string the only field which may acquire an expectation value is the three-form $H$ of the common NS sector 
of string theory. Moreover, this field satisfies a more restrictive Bianchi identity than in the type II case. 
However, apart from the above field, it was suggested that, due to some strong dynamics in the hidden sector,
fermion bilinears may also acquire some vacuum expectation value, thus forming a condensate
 \cite{Dine,Derendinger}. From a Calabi-Yau perspective such condensates are related to supersymmetry 
breakdown. Supersymmetric 
AdS${}_4$ heterotic compactifications on non-K\"ahler manifolds with fluxes and gaugino condensation were 
studied in \cite{LopesCardoso:2003sp,Frey_Lippert}. Moreover, a study including dilatino condensation was 
performed in \cite{Manousselis:2005xa}. However, the above studies do not deal with the solution of the equations 
of motion of the theory but only with the Killing spinor equations and the Bianchi identity. Nevertheless, 
according to Ivanov \cite{ivanov}, it is not straightforward that the solution of the latter imply that the field
equations are satisfied. Therefore it is more natural to directly investigate solutions of the field equations 
of the theory, whether supersymmetric or not. Such a perspective was employed in \cite{Lechtenfeld:2010dr} 
where two such solutions were described, a supersymmetric one with fluxes and gaugino condensation and 
a non-supersymmetric one without gaugino condensation.

In the present paper we extend the analysis of \cite{Lechtenfeld:2010dr} by implementing dilatino condensates 
into the theory. Working at first order in $\alpha'$, we consider the heterotic string including four-fermion 
terms in the action and the supersymmetry variations. Varying this action, the corresponding equations 
of motion are determined. Assuming as background a product of four-dimensional AdS${}_4$ spacetime with
a nearly K\"ahler internal space, all solutions to these equations are derived for two different choices of 
the connection on the tangent bundle. The Bianchi identity is also 
taken into account in order to guarantee the absence of all anomalies. 
An important feature of our investigation, motivated by the results of \cite{ivanov}, is that the gauge 
field is taken to be a generalized instanton. Although it is in principle possible to consider 
non-instanton solutions, gauge fields enjoying the instanton property are distinguished by their 
immediate assurance for the fulfillment of the Yang-Mills equations. 
Apart from the solutions which were 
obtained in \cite{Lechtenfeld:2010dr}, our analysis reveals five sets of new non-supersymmetric solutions. 
Thus a classification of all possible nearly K\"ahler heterotic compactifications with torsion and 
fermion condensates is provided.

The paper is organized as follows. In Section 2 we present the action and the Killing spinor equations 
of the field theory limit of the heterotic string at first order in $\alpha'$, retaining four-fermion 
terms involving the gaugino and the dilatino. Subsequently we determine the field equations resulting from 
this action and their decomposition according to the spacetime factorization AdS${}_4\times K$, where $K$ is
a compact internal space of positive curvature. In Section 3 we briefly describe the geometry of nearly K\"ahler 
manifolds as far as it is needed in our investigation. Section 4 takes $K$ to be one of the four known 
compact homogeneous six-dimensional nearly K\"ahler manifolds and performs
a systematic analysis of possible solutions of the field equations.  We also check which 
of these solutions preserve supersymmetry. Finally, in Section 5 our results are summarized in 
three tables, which present the geometrical and field data as well as the fermion masses for all solutions,
while some interesting properties are discussed as well.

\section{Heterotic strings with fermion condensates}

\paragraph{Field content, action and supersymmetry.\ }
     The low-energy field theory limit of heterotic string theory is given by $d{=}10$, ${\cal N}{=}1$ supergravity coupled to a super-Yang-Mills multiplet and it is defined on the 10$d$ spacetime $M$. The supergravity multiplet consists of  
the graviton $g$, which is a metric on $M$, the left-handed Rarita-Schwinger gravitino $\psi$, the Kalb-Ramond 
two-form field $B$, the scalar dilaton $\phi$ and the right-handed Majorana-Weyl dilatino $\lambda$. 
Moreover, the vector supermultiplet consists of the gauge field one-form $A$ and its superpartner, 
the left-handed Majorana-Weyl gaugino $\chi$. 

Rather than presenting the full action describing the propagation and interactions of the above fields \cite{NIKHEF-H-81-25,NSF-ITP-82-85}, we shall restrict on the part which is relevant for our purposes. 
In this paper we shall consider vacuum solutions where the fermionic expectation values are zero, which is equivalent to the requirement of Lorentz invariance, but certain fermionic bilinears acquire non-trivial vacuum expectation values. However, these vacuum expectation values will not involve the gravitino and therefore it is consistent to set the gravitino to zero from the very beginning,
$\psi = 0$.
Then, in the string frame, the low-energy action up to and including terms of order $\alpha'$
reads as~\cite{BergshdR}~\footnote{Comparing to the action which appears in \cite{BergshdR}, here we have made the 
following field redefinitions: $\phi\rightarrow e^{2\phi/3}$, $\chi\rightarrow\sqrt 2 \chi$ and $H\rightarrow\sfrac 1{3\sqrt 2}H$.}
\begin{eqnarray}\label{hetaction}
\mathcal S(g,\phi,B,\chi,A)=\int_M\!\!d^{10}x\ \sqrt{\det g}\ e^{-2\phi} \!\!\!&\!\!\! \Big\{  
{\text{Scal}} + 4 |d\phi|^2-\sfrac 12|H|^2+\sfrac 12 (H,\Sigma)  - 2 (H,\Delta)+
 \sfrac 14 (\Sigma,\Delta)-\sfrac 18|\Sigma|^2+ \nonumber \\ 
& \ +\ \sfrac14\alpha'\text{tr}\big( 
|\tilde R|^2-|F|^2 -2 \overline\chi \mathcal D \chi-\sfrac 13 \overline\chi\gamma^M\gamma^{AB}F_{AB}\gamma_M\lambda\big)+8\overline\lambda\mathcal D \lambda\Big\},
\nonumber \\ 
\end{eqnarray}
where capital Latin indices run from 0 to 9.
Let us make clear the quantities and the notation in the above action. $\mbox{Scal}$ is the curvature scalar, 
while the curvature forms $F$ and $H$ are 
defined as 
\begin{equation}\label{CurvDefi}
  F\=dA + A\wedge  A \und
  H\=dB + \sfrac14\alpha'\big[\omega_{CS}(\tilde \Gamma)-\omega_{CS}(A)\big],
 \end{equation} 
 where $\omega_{CS}$ denotes the Chern-Simons-forms
 \begin{equation}
 \omega_{CS} (\tilde\Gamma) \= \text{tr} \big(\tilde R \wedge \tilde \Gamma - \sfrac 23\tilde  \Gamma\wedge \tilde \Gamma\wedge \tilde \Gamma\big) \und
 \omega_{CS} (A) \= \text{tr} \big(F \wedge A - \sfrac 23 A\wedge A\wedge A\big),
 \end{equation}
and $\tilde\Gamma$ is a connection on the tangent bundle $TM$, whose choice is ambiguous. This connection 
could be chosen for example to be the Levi-Civita one, $\Gamma^{\text{LC}}(g)$, or a modified connection 
such as the plus or minus ones, $\Gamma^{\pm}=\Gamma^{\text{LC}}\mp\sfrac12H$. We shall return to this point 
later in our analysis. The chosen connection determines the space-time curvature two-form 
\begin{equation}
\tilde R \= d\tilde\Gamma + \tilde\Gamma\wedge\tilde\Gamma.
\end{equation}
Furthermore, in (\ref{hetaction}) appear the following expressions
\begin{equation}
\text{tr}|\tilde R|^2 \= \sfrac 12 \tilde R_{MNPQ}\tilde R^{MNPQ}   \und
\text{tr}|F|^2 \= \sfrac12 \text{tr} F_{MN}F^{MN},
\end{equation}
and traces are taken over the adjoint representation of the gauge group or 
of SO(9,1), depending on the context.
For any two $p$-forms $\alpha,\beta$ we use the definitions 
\begin{equation}
(\alpha,\beta):=\frac{1}{p!}\alpha_{M_1M_2...M_p}\beta^{M_1M_2...M_p}, 
\qquad |\alpha|^2:=(\alpha,\alpha).
\end{equation}
$\mathcal D=\gamma^M\nabla_M$ denotes the Dirac operator, 
coupled to $\Gamma^{\text{LC}}(g)$ and to~$A$. Finally, we have defined
the fermion bilinears
\begin{equation}
 \Sigma \= \sfrac1{24}\alpha'\,
\text{tr} (\overline\chi\gamma_M\gamma_N\gamma_P\chi)\ 
dx^M\wedge dx^N\wedge dx^P
\end{equation}
and
\begin{equation}
 \Delta \= \sfrac1{6}\,
 (\overline\lambda\gamma_M\gamma_N\gamma_P\lambda)\ 
dx^M\wedge dx^N\wedge dx^P.
\end{equation}

The action (\ref{hetaction}) is invariant under ${\cal N}{=}1$ supersymmetry transformations \cite{BergshdR}, which act on the fermions as 
  \begin{align}\label{gauginoSusyVar}
    \delta \psi_M &\= \nabla_M \varepsilon -\sfrac 18H_{MNP}\gamma^N\gamma^P\varepsilon + \sfrac 1{96}\gamma(\Sigma)\gamma_M\varepsilon  , \nonumber\\[4pt]
    \delta \lambda &\= -\sfrac {\sqrt{2}}{4} \gamma\big(d\phi -\sfrac 1{12}H- \sfrac 1{48}\Sigma+\sfrac 1{48}\Delta\big)\varepsilon ,\\[4pt]
    \delta \chi &\= -\sfrac 14 \gamma(F) \varepsilon+\varepsilon\overline\chi\lambda-\chi\overline\varepsilon\lambda
+\gamma^M\lambda\overline\chi\gamma_M\varepsilon, \nonumber
 \end{align}
where $\varepsilon$ is the supersymmetry generator, which is 
a left-handed Majorana-Weyl spinor. In addition,
$\gamma$ denotes the map from forms to the Clifford algebra,
\begin{equation}
 \gamma \big(\sfrac 1{p!}\omega_{M_1\dots M_p}\ dx^{M_1}\wedge\dots\wedge dx^{M_p}\big)\= \omega_{M_1\dots M_p} \gamma^{M_1} \dots\gamma^{M_p}.
\end{equation}

The following additional remarks concerning fermion bilinears are in order. The term in the action which is proportional to $\overline\chi\gamma^M\gamma^{AB}F_{AB}\gamma_M\lambda$ will be ignored in the following. This is a legitimate choice once we assume that there is no 
vacuum expectation value related to bilinears of the mixed form $\overline\chi\lambda$ and the like. Such a choice leads also to a further simplification of the gaugino supersymmetry transformation, which simply becomes
\begin{equation}
\delta \chi \= -\sfrac 14 \gamma(F)\varepsilon.
\end{equation}

\paragraph{Field equations.\ } 

The equations of motion may be obtained by varying the action (\ref{hetaction}), and they take the form (we symmetrize with weight one)
\begin{eqnarray} \label{eom}
\begin{aligned}
\text{Ric}_{MN}  +2 (\nabla d\phi)_{MN} 
-\sfrac 1{8} (H-\sfrac 12\Sigma+2\Delta)_{PQ(M}{H_{N)}}^{PQ}+ \qquad \qquad\qquad\qquad\qquad\qquad\qquad\qquad\\
+\sfrac14\alpha' \Big[ \tilde R_{MPQR}{\tilde R_N}^{\ PQR} 
- \text{tr}\big(F_{MP} {F_N}^P
+\sfrac 12\overline \chi\gamma_{(M} \nabla_{N)}\chi\big)\Big]+2\overline\lambda\gamma_{(M}\nabla_{N)}\lambda&=0, \\
\text{Scal}  - 4 \Delta \phi  +4|d\phi|^2-\sfrac 12|H|^2+\sfrac{1}{2}(H,\Sigma)-2(H,\Delta)+\sfrac 14(\Sigma,\Delta)-\sfrac{1}{8}|\Sigma|^2 + \qquad \qquad\qquad\qquad\quad\\
+\sfrac14\alpha'\text{tr}\Big[ |\tilde R|^2 - |F|^2 - 2\overline\chi\mathcal D\chi\Big]+8\overline\lambda\mathcal D\lambda&=0,\\[2pt]
\big( \mathcal D - \sfrac 1{24} \gamma(H-\sfrac 12\Sigma+\sfrac 12\Delta) \big)\big(e^{-2\phi}\chi\big)&=0 
,\\[4pt]
\big( \mathcal D - \sfrac 1{24} \gamma(H-\sfrac 18\Sigma) \big)\big(e^{-2\phi}\lambda\big)&=0,
 \\[4pt]
e^{2\phi}d\ast (e^{-2\phi}F) + A\wedge\ast F- \ast F\wedge A+\ast (H-\sfrac{1}{2}\Sigma+2\Delta)\wedge F&=0, \\[4pt]
d\ast e^{-2\phi} (H-\sfrac{1}{2}\Sigma+2\Delta) &=0. 
\end{aligned}
\end{eqnarray}
 The derivation of the equations is greatly simplified by a Lemma in \cite{BergshdR}, as was pointed out by Becker and Sethi \cite{BeckerSethi}. It implies that up to this order in $\alpha'$ one can neglect variations of the form $\frac{\partial S}{\partial \tilde \Gamma}\frac {\partial\tilde \Gamma}{\partial(\cdots)}$, for any field~$(\cdots)$.
Apart from the equations of motion, the Bianchi identity for $H$ must be satisfied, which follows from the definition \eqref{CurvDefi}:
 \begin{equation}\label{HBianchi}
  dH\=\sfrac14\alpha' \text{tr}[\tilde R\wedge \tilde R - F\wedge F].
 \end{equation} 

Let us now set the dilaton to a constant value for the remainder of the paper,
\begin{equation}
\phi = \mbox{constant} ,
\end{equation}
which simplifies the equations of motion~(\ref{eom}).
Moreover, taking the trace of the Einstein equation (the first one in~(\ref{eom})) gives
\begin{equation}
  \text{Scal} 
-\sfrac 32 (H-\sfrac{1}{2}\Sigma+2\Delta,H)
   + \sfrac12\alpha' \text{tr}\big[|\tilde R|^2-|F|^2  -\sfrac 12 \overline \chi\mathcal D\chi \big]+4\overline\lambda\mathcal D\lambda \=0. 
\end{equation}
This equation can be combined with 
the dilaton equation (the second one in~(\ref{eom})) in two different ways as
\begin{equation}\label{EinsteinDilatonCombiGaugino}
\begin{aligned}
-|H|^2+\sfrac 14(H,\Sigma)-(H,\Delta)-\sfrac 14(\Sigma,\Delta)+\sfrac 18|\Sigma|^2
+\sfrac14\alpha' \text{tr}\big[ |\tilde R|^2 - |F|^2+\overline\chi\mathcal D\chi\big]-4\overline\lambda\mathcal D\lambda &\=0,\\[4pt]
\text{Scal} 
+\sfrac 12|H|^2+\sfrac 14(H,\Sigma)-(H,\Delta)+\sfrac 12(\Sigma,\Delta)-\sfrac 14|\Sigma|^2
-\sfrac34\alpha' \text{tr}(\overline \chi\mathcal D\chi)+12\overline\lambda\mathcal D\lambda&\= 0.
\end{aligned}
\end{equation}
 One can replace the dilaton equation by one of these, and if further the Einstein equation is satisfied, then the other equation in \eqref{EinsteinDilatonCombiGaugino} is implied. 
Let us note here that setting $\lambda=0$, $\Delta= 0$ one recovers the equations of motion of \cite{Lechtenfeld:2010dr}, as it should be the case.

We would like to make a crucial remark concerning the necessity of writing down and solving all the above 
field equations. It is usually argued that the Killing spinor equations and the Bianchi identity
directly imply the equations of motion. This is indeed the case in type II supergravities, where 
the Bianchi identity is much simpler (see e.g. \cite{grana}). In the heterotic case, one may argue that this 
result is still true at leading order, i.e.\ when $\alpha'$ corrections are ignored. However, crucial 
aspects of the heterotic theory lie in the $\alpha'$ corrections, most importantly the gauge sector in the 
action and the non-trivial corrections in the Bianchi identity. In that case it is far from obvious 
whether the equations of motion follow from the supersymmetry equations and the Bianchi identity \cite{ivanov}. Here 
we do not make such an assumption. Instead we study {\it{all}} the relevant equations independently.

\paragraph{Space-time factorization.\ }
Let us know turn our attention to compactifications of the form
\begin{equation}
M \= \text{AdS}_4(r)\times K,
\end{equation}
with a 4$d$ anti-de Sitter space of `radius'~$r$  
and a 6$d$ compact Riemannian internal space~$K$. 
Lower case Greek indices will be used for the external 4$d$ part, while
lower case Latin indices will be reserved for the internal dimensions.
Furthermore, we assume that $F$, $H$, $\Sigma$ and $\Delta$ are restricted to~$K$, 
i.e.\ they do not depend on the AdS~coordinates. The components of $\tilde R$ 
in AdS direction are taken to coincide with the Riemann curvature of AdS.
This further simplifies the equations.
{}From now on, hatted quantities refer to the AdS~part, 
and unhatted ones live on~$K$. The ambiguity in picking $\tilde\Gamma$
is a choice of connection on~$K$.

In order to properly factorize the spinors, we employ a standard representation
of the $10d$ Clifford algebra
\begin{equation}\label{clifford}
\big\{ \gamma^A\,,\,\gamma^B \big\} \= 2\,\eta^{AB} \= 2\,\text{diag}(-1,+1,\ldots,+1)^{AB} 
\end{equation}
via
\begin{equation}
\big\{\gamma^M\big\} \= \big\{ \widehat\gamma^\mu\otimes\unity_8\ ,\ 
\widehat\gamma_5\otimes\gamma^a \big\} \qquad
\text{for}\qquad M=(\mu,a) \quad \text{with} \quad
\mu=0,1,2,3 \quad\text{and}\quad a=4,\ldots,9 .
\end{equation}
The $10d$, $4d$ and $6d$ chirality operators are 
$\widehat\gamma_5{\otimes}\gamma$, 
$\widehat\gamma_5=\widehat\gamma_0\widehat\gamma_1\widehat\gamma_2\widehat\gamma_3$ and
$\gamma=\gamma_4\gamma_5\gamma_6\gamma_7\gamma_8\gamma_9$, respectively.
The gaugino is taken to factorize as
\begin{equation} \label{chifactor}
\chi \= e^{\frac{i\pi}4}\,\widehat\chi\otimes\eta + e^{-\frac{i\pi}4}\,\widehat\chi^*\otimes\eta^*,
\end{equation}
where $\widehat\chi$ is an anticommuting positive-chirality Weyl spinor on $\text{AdS}_4$
with values in the adjoint of the gauge group, while $\eta$ denotes a commuting positive-chirality
Weyl spinor on~$K$, 
\begin{equation}
\widehat\gamma_5\widehat\chi=\widehat\chi, \qquad
\widehat\gamma_5\widehat\chi^\ast=-\widehat\chi^\ast, \qquad
\gamma\,\eta=\eta, \qquad \gamma\,\eta^\ast=-\eta^\ast,
\end{equation}
and we assume $\eta$ to be normalized: $\overline \eta \eta=1$.\footnote{
Note that $\widehat{\overline\chi}=\widehat{\chi}^\dagger\gamma^0$ but $\overline\eta=\eta^\dagger$.}
Similarly, we decompose the dilatino as
\begin{equation}\label{lambdafactor}
\lambda=e^{\frac{i\pi}4}\,\widehat\lambda\otimes\eta^{\ast}+e^{-\frac{i\pi}4}\,\widehat\lambda^{\ast}\otimes\eta,
\end{equation}
with
\begin{equation}
\widehat\gamma_5\widehat\lambda=\widehat\lambda, \qquad
\widehat\gamma_5\widehat\lambda^\ast=-\widehat\lambda^\ast,
\end{equation}
keeping in mind that its chirality is opposite to that of the gaugino.

As a consequence of the splitting, the equations of motion (\ref{eom}) 
decompose. We suppress the tensor product symbol.
The Einstein equation (first in (\ref{eom})) splits into 
\begin{eqnarray} \label{Einsteinsplit}
\widehat{\text{Ric}}_{\mu\nu} +\sfrac14\alpha' 
\widehat{R}_{\mu\alpha\beta\gamma}{\widehat{R}_\nu}^{\ \alpha\beta\gamma} \!\!&=&\!\!
\sfrac18\alpha'\text{tr} \big(\overline \chi \widehat\gamma_{(\mu}\widehat\nabla_{\nu)}\chi\big)
-2\overline\lambda\widehat\gamma_{(\mu}\widehat\nabla_{\nu)}\lambda , \nonumber\\[4pt]
 \sfrac 18\alpha' \text{tr}\big(\overline\chi \big(\widehat \gamma_\mu \nabla_a + \widehat\gamma_5\gamma_a\widehat \nabla_\mu\big)\chi\big) \!\!&=&\!\!
 2\overline\lambda\big(\widehat\gamma_{\mu}\nabla_a+\widehat\gamma_5\gamma_a\widehat\nabla_{\mu}\big)\lambda, \nonumber\\[4pt]
\text{Ric}_{ab} -\sfrac 18 (H-\sfrac 12 \Sigma+2\Delta)_{cd(a}{H_{b)}}^{cd}
+\sfrac14\alpha' \big[ \tilde R_{acde}{\tilde R_b}^{\ cde} 
- \text{tr}(F_{ac} {F_b}^c)\big] \!\!&=&\!\!
\sfrac18\alpha'\text{tr}\big(\overline \chi \widehat\gamma_5\gamma_{(a}\nabla_{b)}\chi\big)-2\overline\lambda\widehat\gamma_5\gamma_{(a}\nabla_{b)}\lambda, \nonumber\\
\end{eqnarray}
while the two combinations~(\ref{EinsteinDilatonCombiGaugino}) take the form
\begin{equation}\label{combinations1}
\begin{aligned}
 & -(H-\sfrac 14\Sigma+\Delta,H)+\sfrac 18(\Sigma-2\Delta,\Sigma)
+\sfrac14\alpha' \text{tr}\big[ |\widehat{R}|^2 + |\tilde R|^2 - |F|^2 \big]\= 
-\sfrac14\alpha' \text{tr}\big({\overline\chi}
(\widehat{\mathcal D}{+}\mathcal D)\chi \big)+4\overline\lambda(\widehat{\mathcal D}{+}\mathcal D)\lambda, \\[4pt]
& \widehat{\text{Scal}} + \text{Scal} 
+\sfrac 12(H+\sfrac 12\Sigma-2\Delta,H)-\sfrac 14(\Sigma-2\Delta,\Sigma)
 \=  \sfrac34\alpha' \text{tr}
\big({\overline\chi}
(\widehat{\mathcal D}{+}\mathcal D)\chi\big)-12\overline\lambda(\widehat{\mathcal D}{+}\mathcal D)\lambda .
\end{aligned}
\end{equation}
 The gaugino and dilatino equations become
\begin{equation} \label{gauginodilatino}
\begin{aligned}
\big(\widehat{\mathcal D}+\mathcal D-\sfrac1{24} \gamma(H-\sfrac 12\Sigma+\sfrac 12\Delta)\big)\ \chi&\=0, 
\\[4pt]
\big(\widehat{\mathcal D}+\mathcal D-\sfrac1{24} \gamma(H-\sfrac 18\Sigma) \big)\ \lambda &\=0,
\end{aligned}
\end{equation}
which straightforwardly yields
\begin{eqnarray}
\alpha'\,\text{tr} \big({\overline\chi}
(\widehat{\mathcal D}{+}\mathcal D)\chi\big)\!&=&\!(H-\sfrac 12 \Sigma+\sfrac 12 \Delta\,,\,\Sigma),
\\[4pt] 
\overline\lambda(\widehat{\mathcal D}{+}\mathcal D)\lambda\!&=&\!\sfrac 14 (H-\sfrac 18\Sigma\,,\,\Delta)
\end{eqnarray}
for the fermion kinetic terms, further simplifying (\ref{combinations1}).

Moreover, the gravitational data on $\text{AdS}_4(r)$ are
\begin{equation}
\widehat{\text{Scal}}=-\frac {12}{r^2}, \qquad
\widehat{\text{Ric}}= \sfrac14\widehat{\text{Scal}}\ \widehat{g} =
-\frac{3}{r^2}\,\widehat{g}, \qquad
\widehat{R}_{\mu\alpha\beta\gamma}{\widehat{R}_\nu}^{\alpha\beta\gamma} = 
\sfrac1{24}\widehat{\text{Scal}}^2 \widehat{g}_{\mu\nu} 
=\frac {6}{r^4}\, \widehat{g}_{\mu\nu}.
\end{equation}

Inserting these relations into the equations of motion, we obtain
\begin{equation}\label{EOM_split}
\begin{aligned}
& -\big(\sfrac3{r^2}-\sfrac3{2r^4}\alpha'\big)\,\widehat{g}_{\mu\nu}
\=\sfrac18\alpha'\text{tr}\big(
\widehat{\overline\chi}\widehat\gamma_{(\mu}\widehat\nabla_{\nu)}\widehat\chi + \widehat{\overline\chi}^*\widehat\gamma_{(\mu}\widehat\nabla_{\nu)}\widehat\chi ^*
\big)-2\big(
\widehat{\overline\lambda}\widehat\gamma_{(\mu}\widehat\nabla_{\nu)}\widehat\lambda + \widehat{\overline\lambda}^*\widehat\gamma_{(\mu}\widehat\nabla_{\nu)}\widehat\lambda ^*
\big),\\[6pt]
& \sfrac18\alpha'\text{tr}\big( (\widehat{\overline\chi}\widehat\gamma_\mu\widehat\chi)
(\overline\eta\nabla_a\eta)+(\widehat{\overline{\chi}}^\ast \widehat\gamma_\mu \widehat\chi^\ast)(\overline\eta^\ast\nabla_a\eta^\ast) - 
(\widehat{\overline\chi}\widehat\gamma_5\widehat\nabla_\mu\widehat\chi^\ast)(\overline\eta\gamma_a\eta^\ast) +(\widehat{\overline\chi}^\ast\widehat\gamma_5\widehat\nabla_\mu\widehat\chi)(\overline\eta^\ast\gamma_a\eta)\big)\=\\ 
& \qquad\qquad\qquad \= 2\big( (\widehat{\overline\lambda}\widehat\gamma_\mu\widehat\lambda)
(\overline\eta^\ast\nabla_a\eta^\ast)+(\widehat{\overline{\lambda}}^\ast \widehat\gamma_\mu \widehat\lambda^\ast)(\overline\eta\nabla_a\eta) - 
(\widehat{\overline\lambda}\widehat\nabla_\mu\widehat\lambda^\ast)(\overline\eta^\ast\gamma_a\eta) +(\widehat{\overline\lambda}^\ast \widehat\nabla_\mu \widehat \lambda)(\overline\eta\gamma_a\eta^\ast)\big),\\[6pt]
& \text{Ric}_{ab} -\sfrac 18 (H-\sfrac 12\Sigma+2\Delta)_{cd(a}{H_{b)}}^{cd}
+\sfrac14\alpha' \big[ \tilde R_{acde}{\tilde R_b}^{\ cde} 
- \text{tr}(F_{ac} {F_b}^c)\big]\= \\
& \= \sfrac{i}8\alpha'\text{tr}\big( 
(\widehat{\overline\chi}^\ast\widehat\gamma_5\widehat\chi)
(\overline\eta^\ast\gamma_{(a}\nabla_{b)}\eta)+
(\widehat{\overline\chi}\widehat\gamma_5\widehat\chi^\ast)
(\overline\eta\gamma_{(a}\nabla_{b)}\eta^\ast) \big)
-2i\big( 
(\widehat{\overline\lambda}^\ast\widehat\gamma_5\widehat\lambda)
(\overline\eta\gamma_{(a}\nabla_{b)}\eta^\ast)+
(\widehat{\overline\lambda}\widehat\gamma_5\widehat\lambda^\ast)
(\overline\eta^\ast\gamma_{(a}\nabla_{b)}\eta) \big),\\[6pt]
& -(H-\sfrac 12\Sigma+2\Delta\,,\,H)
+\sfrac3{r^4}\alpha'+\sfrac14\alpha' \text{tr}\big[|\tilde R|^2 - |F|^2 \big]\=0,\\[4pt]
&  -\sfrac{12}{r^2} + \text{Scal}
+\sfrac 12(H-\Sigma+4\Delta\,,\,H)+\sfrac 18(\Sigma-2\Delta\,,\,\Sigma)\= 0, \\[4pt]
& \big(\widehat{\mathcal D}+\mathcal D-\sfrac1{24} \gamma(H-\sfrac 12\Sigma+\sfrac 12\Delta)\big)\ \chi\=0, 
\\[4pt]
& \big(\widehat{\mathcal D}+\mathcal D-\sfrac1{24} \gamma(H-\sfrac 18\Sigma) \big)\ \lambda\=0, \\[4pt]
& d\ast F + A\wedge\ast F- \ast F\wedge A+\ast (H-\sfrac{1}{2}\Sigma+2\Delta)\wedge F\=0, \\[4pt]
& d\ast (H-\sfrac{1}{2}\Sigma+2\Delta) \=0.
\end{aligned}
\end{equation}
These equations entangle the internal fields with the AdS data. 

On the right-hand side of these equations we encounter external gaugino and dilatino
bilinears, which are nilpotent on the classical level. The standard lore 
to give meaning to these terms performs a quantum average $\langle\dots\rangle$
over the space-time fermionic degrees of freedom.
At this stage, assumptions about the fermionic quantum correlators enter:
we assume the presence of a suitable space-time gaugino condensate as 
a backdrop for the bosonic equations, namely
\begin{equation} \label{condensate}
\langle\tr\,\widehat{\overline\chi}\widehat\gamma_5\widehat\chi^\ast\rangle \= i\,\Lambda^3
\qquad\text{but}\qquad
\langle\tr\,\widehat{\overline\chi}\widehat{M}\widehat\chi \rangle \=
\langle\tr\,\widehat{\overline\chi}\widehat{M}\widehat\chi^\ast \rangle \= 0
\end{equation}
for all non-scalar operators $\widehat{M}$.
The condensate scale~$\Lambda\in\mathbb R$ will be fixed later. 
Similar considerations hold for the dilatino condensate, i.e.
\begin{equation}\label{condensate2}
 \langle\,\widehat{\overline\lambda}\widehat\gamma_5\widehat\lambda^\ast\rangle 
 \= i\,\tilde\Lambda^3
\qquad\text{but}\qquad
\langle\,\widehat{\overline\lambda}\widehat{M}\widehat\lambda \rangle \=
\langle\,\widehat{\overline\lambda}\widehat{M}\widehat\lambda^\ast \rangle \= 0.
\end{equation}

After averaging over the gaugino and dilatino,
our set of equations (\ref{EOM_split}) simplify to
\begin{equation} \label{EOM_condensate}
\begin{aligned}
 -\big(\sfrac3{r^2}-\sfrac3{2r^4}\alpha'\big)\,\widehat{g}_{\mu\nu} &\=0,\\[4pt]
\text{Ric}_{ab} -\sfrac 18 (H-\sfrac 12\Sigma+2\Delta)_{cd(a}{H_{b)}}^{cd}
+\sfrac14\alpha' \big[ \tilde R_{acde}{\tilde R_b}^{\ cde} 
- \text{tr}(F_{ac} {F_b}^c)\big] &\= \\
\= (\sfrac18\alpha'\Lambda^3+2\tilde{\Lambda}^3)&\big(
\overline\eta^\ast\gamma_{(a} \nabla_{b)}\eta-
\overline\eta\,\gamma_{(a} \nabla_{b)}\eta^\ast \big),\\[4pt]
-(H-\sfrac 12\Sigma+2\Delta\,,\,H)
+\sfrac3{r^4}\alpha'+\sfrac14\alpha' \text{tr}\big[|\tilde R|^2 - |F|^2 \big]&\=0, \\[4pt]
  -\sfrac{12}{r^2} + \text{Scal}
+\sfrac 12(H-\Sigma+4\Delta\,,\,H)+\sfrac 18(\Sigma-2\Delta\,,\,\Sigma)&\= 0, \\[4pt]
 \big(\widehat{\mathcal D}+\mathcal D-\sfrac1{24} \gamma(H-\sfrac 12\Sigma+\sfrac 12\Delta)\big)\ \chi&\=0, 
 \\[4pt]
 \big(\widehat{\mathcal D}+\mathcal D-\sfrac1{24} \gamma(H-\sfrac 18\Sigma) \big)\ \lambda&\=0, \\[4pt]
 d\ast F + A\wedge\ast F- \ast F\wedge A+\ast (H-\sfrac{1}{2}\Sigma+2\Delta)\wedge F&\=0, \\[4pt]
 d\ast (H-\sfrac{1}{2}\Sigma+2\Delta) &\=0,
\end{aligned}
\end{equation}
where we continue to use the symbol $\Sigma$ for the condensate
\begin{equation}
\langle \Sigma \rangle \= \sfrac{1}{24}\,\Lambda^3\alpha'\big(
\overline\eta^\ast\gamma_a\gamma_b\gamma_c\eta +
\overline\eta\,\gamma_a\gamma_b\gamma_c\eta^\ast \big)\,
dx^a\wedge dx^b\wedge dx^c 
\end{equation}
and the symbol $\Delta$ for 
\begin{equation}
 \langle\Delta\rangle \= \sfrac{1}{6}\,\tilde\Lambda^3\big(
\overline\eta^\ast\gamma_a\gamma_b\gamma_c\eta +
\overline\eta\,\gamma_a\gamma_b\gamma_c\eta^\ast \big)\,
dx^a\wedge dx^b\wedge dx^c. 
\end{equation}

Remarkably,
the first equation fixes the $\text{AdS}_4$ radius in terms of~$\alpha'$,
\begin{equation}
r^2 \= \sfrac12\alpha'.
\end{equation}
So the $\alpha'$ corrections to heterotic supergravity are essential for obtaining 
an $\text{AdS}$ solution,\footnote{
This was already observed in~\cite{Green} who determined the sign of the 
cosmological constant in the absence of fermion condensates.} 
but in our framework the Einstein equations do not admit a $\text{dS}$~spacetime.
In the fourth equation (the dilaton equation),
the negative contribution $\widehat{\text{Scal}}=-\frac{12}{r^2}$ 
allows for internal manifolds of positive scalar curvature, 
which are excluded in Minkowski compactifications. A certain class of such internal 
spaces is given by nearly K\"ahler manifolds, which we study in the present paper.

\section{Geometry of nearly K\"ahler manifolds}

In the present section we briefly describe the basics on the geometry of nearly K\"ahler manifolds in order 
to collect the necessary ingredients for the ensuing analysis in the next section. None of this material 
is new, and a more detailed account may be found in \cite{Lechtenfeld:2010dr}.

Nearly K\"ahler manifolds comprise a subclass of SU(3) structure manifolds, i.e.\ manifolds possessing a 
nowhere-vanishing, globally defined spinor $\eta$ which is covariantly constant with respect to a 
connection with torsion.\footnote{
The same notation $\eta$ for the spinor as in (\ref{chifactor}) is used 
since the two spinors will be identified.} 
Manifolds with SU(3) structure constitute a broader class than 
Calabi-Yau manifolds, namely manifolds with SU(3) holonomy. Indeed, Calabi-Yau manifolds are included in the 
former, and they correspond to the case when the torsion vanishes and the connection reduces to the Levi-Civita 
one. 

Based on the spinor $\eta$, the structure forms of a nearly K\"ahler manifold can be constructed. They are a 
real two-form $\omega$ of type $(1,1)$ and a complex three-form $\Omega$ of type $(3,0)$, given as
  \begin{align}
\omega&\=\sfrac i2\,\overline\eta\,\gamma_a\gamma_b\eta\,e^a\wedge e^b,
\nonumber\\[4pt]
\Omega&\=+\sfrac 16\,\overline\eta\,\gamma_a\gamma_b\gamma_c\eta^\ast
e^a\wedge e^b\wedge e^c, \\[4pt]
\overline\Omega&\=-\sfrac 16\,\overline\eta^\ast\gamma_a\gamma_b\gamma_c\eta\,
e^a\wedge e^b\wedge e^c, \nonumber
\end{align}
where the $e^a$ form an orthonormal frame of one-forms for $T^\ast(K)$. The structure forms are not closed, 
which would be the case for a Calabi-Yau manifold, but instead they satisfy the conditions
\begin{equation}\label{NK_defRels}
d\omega \= -3\,\varsigma\, \text{Re}(\Omega),\qquad 
d\Omega \= 2i\varsigma\, \omega\wedge \omega
\qquad\text{with}\quad \varsigma\in\mathbb R.
\end{equation} 
Moreover, the above forms obey the duality relations
\begin{equation} \label{duality}
{\ast}\Omega \= -i\Omega,\qquad 
\ast\overline\Omega \= i\overline \Omega, \qquad
2\,{\ast}\omega \= \omega\wedge \omega,
\end{equation}
and they act on the spinors $\eta$ and $\eta^\ast$ as 
\begin{equation}\label{OmegaEtaProp}
 \begin{aligned}
   \Omega_{abc}\gamma^c \eta^\ast&\=0, \qquad\qquad\qquad
   \overline\Omega_{abc}\gamma^c \eta\=0, \\[4pt]
   \Omega_{abc}\gamma^b\gamma^c\eta &\= -8\gamma_a \eta^\ast,\qquad 
   \overline\Omega_{abc}\gamma^b\gamma^c\eta^\ast \= 8\gamma_a \eta.
 \end{aligned}
\end{equation}
Their normalization is
\begin{equation}
(\omega,\omega) = 3,\qquad 
(\Omega,\overline\Omega) =8,\qquad 
(\Omega,\Omega)= (\overline\Omega,\overline\Omega)=0,
\end{equation}
where $(\cdot,\cdot)$ denotes the metric induced on $\Omega(K)$ by $g$. 
This implies
\begin{equation}\label{NK_Form-Vol-Rel}
\omega^3=6\,\text{Vol} \und \Omega\wedge \overline \Omega = -8i\,\text{Vol}.
\end{equation} 

An important quantity in the study of a nearly K\"ahler manifold is its intrinsic torsion. 
This is defined as the torsion of the canonical connection. 
It is totally antisymmetric and it is proportional to the imaginary part 
of the structure three-form, $\text{Im}(\Omega)$. 
Here we identify the $H$-flux with the intrinsic torsion, 
\begin{equation}\label{H-OmegaRel}
H\= -\sfrac{i}2\varsigma\,\big(\Omega-\overline \Omega\big)\=
\varsigma\,\text{Im}(\Omega).
\end{equation}
Therefore, the canonical connection reads
\begin{equation}
\nabla^- \= \nabla^{LC}\ +\ \sfrac12 H^a_{bc}\,e^b \otimes (E_a\otimes e^c),
\end{equation}
with vector fields $E_a$ dual to the one-forms $e^a$. 

According to \cite{butruille}, there exist four homogeneous six-dimensional nearly K\"ahler manifolds, which 
can be represented as coset spaces $K=G/H$, for two Lie groups $H\subset G$, where $H$ is isomorphic to a 
subgroup of SU(3), and not to be confused with the torsion form $H\in \Omega(K)$:
\begin{equation}
\begin{aligned}
\text{SU(3)}/\text{U(1)}{\times}\text{U(1)},\qquad & \quad 
\text{Sp(2)}/\text{Sp(1)}{\times}\text{U(1)},\\[2pt]
G_2/\text{SU(3)}=S^6,\quad\qquad & 
\text{SU(2)}^3/\text{SU(2)}_{\text{diag}}=S^3{\times}S^3.
\end{aligned}
\end{equation}

Here we shall not delve into details regarding the geometry of the above coset spaces; 
the reader may consult \cite{Castellani,Mullerhoissen}. However, let us simply collect 
the results on connections and other relevant quantities, which are necessary for the 
analysis in the next section. According to the coset structure $G/H$, we decompose the
Lie algebra of $G$ as ${\mathfrak g}={\mathfrak h}\oplus{\mathfrak m}$.
A $G$-invariant metric on the coset spaces is given as
\begin{equation}
 g _{ab} \=- f^c_{ad}f^d_{bc} - 2f^c_{ak}f^k_{bc},
\end{equation} while the Levi-Civita connection  is
\begin{equation}
   \Gamma \= \big(f^a_{ic}e^i+\sfrac12 f^a_{bc}e^ b\big)\otimes(E_a \otimes e^ c) ,
 \end{equation}  
where $\{E_k\}_{k=7,\dots,\dim{\mathfrak g}}$ is a basis of the Lie algebra $\mathfrak h$ of $H$, $\{E_a\}_{a=1,\dots ,6}$ is a basis of the coset part $\mathfrak m$, 
and $\{e^a\}$, $\{e^k\}$ are the dual bases. 
As indicated, we use letters $a,b,c,\dots$ for indices in $\mathfrak m$ and $i,j,k,\dots$ for those in $\mathfrak h$. As usual, $f$s denote the Lie algebra structure constants. 
One may define a family of metric connections $\nabla^\kappa$ whose torsion is
\begin{equation}
T\=\kappa\,f^a_{bc}\,e^b\otimes (E_a\otimes e^c)\ \in\Omega^1(\text{End}(TK))
\qquad\text{with}\quad\kappa\in\mathbb R. 
\end{equation}
The canonical connection is included for $\kappa=-1/2$. Thus we have
\begin{equation}\label{kappaConnection}
\Gamma^\kappa \= 
\big(f^a_{ic}e^i+\sfrac12 \tau\,f^a_{bc}e^ b\big)\otimes(E_a \otimes e^ c) 
\qquad\text{with}\quad \tau:=2\kappa{+}1.
\end{equation} 
In the following we will denote $\nabla^\kappa$ for special values 
of $\kappa$ as $\nabla^{-\frac12}=:\nabla^-$, $\nabla^{0}=:\nabla\ $ and $\nabla^{+\frac12}=:\nabla^+$.
We note that the torsion three-form of the canonical connection $\nabla^-$ reads
\begin{equation}
H\= -\sfrac 16 f_{abc}\,e^a\wedge e^b\wedge e^c.
\end{equation} 
Based on the above formulae the curvature tensors for the nearly K\"ahler manifolds may be computed in terms 
of their structure constants. The results appear in \cite{Lechtenfeld:2010dr}; here we just list a set 
of curvature related quantities which appear in the equations of motion (\ref{EOM_condensate}) 
to be solved:
\begin{equation}\label{BianchiExpressions}
\begin{aligned}
& \text{tr}_\mathfrak m R^\kappa\!\wedge R^\kappa \= (\beta{+}\sfrac14\tau^2{-}1)\,dH
\qquad\Longrightarrow\qquad
\text{tr}_\mathfrak m R^+\!\wedge R^+ \= \beta\, dH, \qquad\ \,
\text{tr}_\mathfrak m R^-\!\wedge R^- \= (\beta{-}1)\,dH, \\
& \text{tr}_\mathfrak m |R^\kappa|^2 \= \sfrac1{24}\tau^2\big(\tau^2{-}2\big) +1-\beta
\quad\,\Longrightarrow\qquad
\text{tr}_\mathfrak m |R^+|^2 \= \sfrac 43{-}\beta, \qquad\qquad
\text{tr}_\mathfrak m |R^-|^2 \= 1{-}\beta, \\
& \text{tr}_\mathfrak h R^-\!\wedge R^- \= -\beta\, dH, \qquad\ \
\text{tr}_\mathfrak h |R^-|^2 \= \beta, \\
& R^+_{acde}R_{\, b}^{+cde} \= \sfrac {4-3\beta}9\,g_{ab}, \qquad
R^-_{acde}R_{\, b}^{-cde} \= \sfrac {1-\beta}3\,g_{ab},
\end{aligned}
\end{equation}
where $\beta$ is a constant, characteristic of each manifold, with value
\begin{center}
\begin{equation}\label{betas}
\begin{tabular}{|c|c|c|c|c|}
\hline
&SU(3)/U(1)$\times$U(1)&Sp(2)/Sp(1)$\times$U(1)&$G_2$/SU(3)&SU(2)$^3$/SU(2) \\ 
\hline 
$\beta$ & 0 & -- & 3/4 & 1/3 \\
\hline
\end{tabular}
\end{equation}
\end{center}

In the compactifications we consider, 
the curvature $\tilde R$ of the tangent bundle of $K$ is one of the $R^\kappa$,
usually either $R^+$ or $R^-$. In contrast, the gauge field $F$ is free to live
on an arbitrary bundle, and so we also have the choice $F=R^-\big|_\mathfrak h$ 
at our disposal. The supersymmetry constraint, however, forces $F$ to be a 
(generalized) instanton, meaning that $\ast F= -\omega\wedge F$. 
This is satisfied only by $F=R^-$ (both on $\mathfrak m$ and on $\mathfrak h$). 
If $\mathfrak h$ is abelian, we also have the freedom to rescale 
$\Gamma^-$ without losing the instanton property. Hence, for 
SU(3)/U(1)$\times$U(1) we may take $F=\lambda R^-$ with $\lambda\in\mathbb R$ 
(cf.\ \cite{Popov09}). Yet even without the supersymmetry constraint it is very
convenient to choose an instanton solution for the gauge field, because it 
automatically satisfies the Yang-Mills equation. 

On the space Sp(2)/Sp(1)$\times $U(1) there is no common value for $\beta$. 
Instead, we have $\beta=0$ on $\mathfrak u(1)$ and $\beta = 2/3$ on $\mathfrak{sp}(1)$. 
We shall return to this point in the following section where we explain 
how this case should be treated.

Finally, we can compare the general theory to our concrete realization 
in terms of coset models. In particular, we identified
\begin{equation}
T^c_{ab}=-\sfrac 12f^c_{ab} \und \text{Scal}\equiv\text{Scal}^0=\sfrac52 
\qquad\Longrightarrow\qquad
\varsigma \= \sqrt{\sfrac{\text{Scal}}{30}} \= \sqrt{\sfrac 1{12}}.
\end{equation}
{}From this we easily deduce
\begin{equation}\label{fHImOmega}
(\Omega-\overline\Omega)_{abc}\=-4\sqrt 3\,i\,f_{abc} \und H_{abc}\=-f_{abc}
\end{equation}
as well as the relations
\begin{align}
  d\omega &\= -\sfrac{3}2\varsigma\,\big(\Omega +\overline \Omega\big) \=-\sfrac{\sqrt 3}4\big(\Omega +\overline \Omega\big)\= 3\,{\ast}H,\nonumber\\[2pt]
   d\Omega &\=-d\overline \Omega \=2i\varsigma\,\omega\wedge \omega \= \sfrac 1{\sqrt 3} i\,\omega\wedge \omega, \\[2pt]
  dH &\= \sfrac1{15} \text{Scal}\,\omega\wedge \omega \= \sfrac 16\,\omega\wedge \omega.\nonumber
 \end{align}

With $\varsigma$ we have fixed the scale-squared $\rho=\frac{1}{12\varsigma^2}$
of the nearly K\"ahler manifold to be unity.
To employ such manifolds for heterotic string solutions, 
we must reintroduce a dimensionful scale $\sqrt{\rho}$ by
replacing $\varsigma^2\to\sfrac1{12\rho}$.
Dimensional analysis then yields $g\to\rho g$ for the metric and
\begin{equation}
\begin{aligned}
\text{Ric}&\=\sfrac 5{12}\rho^{-1} g,\qquad\,
\text{Scal}\=\sfrac 5{2}\rho^{-1},\\[2pt]
|H|^2 &\= \sfrac13 \rho^{-1},\qquad\quad 
H_{acd}{H_b}^{cd} \=\sfrac13 \rho^{-1} 
g_{ab}, 
\end{aligned}
\end{equation}
and the quantities given in \eqref{BianchiExpressions} scale as follows,
\begin{equation}
\text{tr } R\wedge R\ \sim\ \rho^{-1} dH, \qquad 
\text{tr } |R|^2\ \sim\ \rho^{-2},\qquad 
R_{acde}{R_b}^{cde}\ \sim\ \rho^{-2}g_{ab}.
\end{equation}

\section{Solutions with gaugino and dilatino condensates}

In order to find a supersymmetric solution of the heterotic string on nearly K\"{a}hler manifolds,
we have to guarantee that all the equations of motion are satisfied, 
that the supersymmetry variations vanish and that the Bianchi identity holds. Therefore, in the supersymmetric case 
the full set of equations (\ref{EOM_condensate}), (\ref{gauginoSusyVar}) and (\ref{HBianchi}) has to be solved. 
In the case of non-supersymmetric solutions, the Killing spinor 
equations do not have to be satisfied. 

The strategy we shall follow amounts to the following steps. First, 
since $H_{abc}$ is proportional to $f_{abc}$, it is natural 
to demand that the condensates $\Sigma$ and $\Delta$ be also proportional 
to the structure constants.  Let us implement this feature by writing
\begin{equation}\label{mndef}
 \Sigma = mH \qquad\text{and} \qquad \Delta=nH,
\end{equation}
where $m$ and $n$ are real constants which will be determined from consistency with 
the equations of motion. Second, we will insert these relations into the conditions
for supersymmetry and, third, into the equations of motion and the Bianchi identity.
Fourth, the resulting equations will be scanned for solutions, for the choices of
$\tilde\Gamma=\Gamma^-$ and $\tilde\Gamma=\Gamma^+$. 
Fifth, the remaining equations of motion will be checked for all candidate solutions.

\paragraph{Conditions for supersymmetry.} 

The supersymmetry generator $\varepsilon$ appearing in (\ref{gauginoSusyVar}) 
is obtained as follows 
\cite{Frey_Lippert}: AdS$_4(r)$ carries a Killing spinor 
$\widehat\zeta+\widehat\zeta^\ast$ with Killing number 
$\vartheta = \frac 1{2r}=\frac 1{\sqrt{2\alpha'}}$ \cite{Baum}, i.e.
\begin{equation}\label{AdSKillingSpinor}
\nabla_\mu \widehat\zeta \= \vartheta\,\gamma_\mu\widehat\zeta^\ast \und
\nabla_\mu \widehat\zeta^\ast \= \vartheta\,\gamma_\mu\widehat\zeta.
\end{equation} 
On $K$ we have the positive-chirality Killing spinor $\eta$ with $\nabla^- \eta=\nabla^-\eta^\ast=0$, which gives
\begin{equation}\label{10dSpinor}
\varepsilon \=e^{\frac{i\pi}4}\,\widehat\zeta\otimes\eta
\ +\ e^{-\frac{i\pi}4}\,\widehat\zeta^\ast \otimes \eta^\ast.
\end{equation} 

Let us first study the dilatino supersymmetry variation. 
Since the dilaton has a constant value, it is straightforward to see that
\begin{equation}
 \delta\lambda=0 \quad  \Leftrightarrow \quad H = -\sfrac 14(\Sigma+\Delta). 
\end{equation}
This equation may be rewritten in terms of the constants $m$ and $n$ as
\begin{equation}\label{c1}
 m+n+4=0,
\end{equation}
and thus it provides a first condition on the two coefficients.

Turning to the gravitino variation, for the AdS$_4$ components of $\delta\psi_\mu$ we obtain
\begin{equation}
\begin{aligned}
\delta\psi_\mu &\= \nabla_\mu \varepsilon\ +\ \sfrac 1{96}\gamma(\Sigma)\gamma_\mu \varepsilon \\[2pt]
      &\=e^{\frac{i\pi}4} \vartheta\,\gamma_\mu \widehat\zeta^\ast \otimes \eta\ +\ e^{-\frac{i\pi}4}\vartheta\,\gamma_\mu\widehat\zeta \otimes \eta^\ast\ -\ \sfrac 1{96}\gamma_\mu\gamma(\Sigma)\big(e^{\frac{i\pi}4}\widehat\zeta \otimes \eta + e^{-\frac{i\pi}4}\widehat\zeta^\ast \otimes \eta^\ast\big).
\end{aligned}
\end{equation}
From (\ref{H-OmegaRel}) and (\ref{fHImOmega}), it follows that
\begin{equation}
\gamma(H) \= -\sfrac{i}{2}\,\varsigma\,\gamma(\Omega-\overline\Omega) 
\= -\sfrac{1}{\sqrt{\rho}}\,f_{abc}\,\widehat\gamma_5\,\gamma^a\gamma^b\gamma^c,
\end{equation}
which together with (\ref{OmegaEtaProp}) and (\ref{mndef}) implies
\begin{equation}
\gamma(\Sigma)\,\widehat\zeta{\otimes}\eta \= 24mi\,\varsigma\,\widehat\zeta{\otimes}\eta^\ast \und
\gamma(\Sigma)\,\widehat\zeta^\ast{\otimes}\eta^\ast \=24mi\,\varsigma\,\widehat\zeta^\ast{\otimes}\eta.
\end{equation}
Finally we get
\begin{equation}\label{c2}
0\= \delta \psi_\mu \= (\vartheta+\sfrac{m}{4}\varsigma)
\Big[e^{\frac{i\pi}4}\gamma_\mu\,\widehat\zeta^\ast\!\otimes\eta + 
e^{-\frac{i\pi}4}\gamma_\mu\,\widehat\zeta\otimes \eta^\ast\Big]
\qquad\Longrightarrow\qquad \vartheta=-\sfrac m4\varsigma
\qquad\Longrightarrow\qquad \rho=\sfrac{m^2}{96}\alpha',
\end{equation} 
hence the vanishing of the external gravitino variation fixes the internal scale as well.
The internal gravitino variation $\delta \psi_a$ gives zero anyway, 
due to $\gamma(\Sigma)\gamma_a\eta=\gamma(\Sigma)\gamma_a\eta^\ast=0$ \cite{Lechtenfeld:2010dr}. 

Concerning the last Killing spinor equation, the gaugino variation vanishes 
when the gauge field is a generalized instanton. 
Therefore, we conclude that the conditions (\ref{c1}) and (\ref{c2}) and the instanton property 
for the gauge field suffice to guarantee that the three Killing spinor equations are satisfied,
thus that supersymmetry is intact.

\paragraph{Conditions from the equations of motion and the Bianchi identity.}

Let us turn our attention to the equations of motion and the Bianchi identity. 
We begin with the fourth and third equations in (\ref{EOM_condensate}) and the Bianchi identity 
(\ref{HBianchi}), since these will already severely constrain the available parameter space. 
Let us write the above equations in terms of $\rho,m,n$. 
The fourth equation of (\ref{EOM_condensate}) yields
\begin{equation}\label{eom4}
\frac{12}{r^2}+\frac{5}{2\rho}\= \frac{1}{6\rho}\,\big(\sfrac14 m^2{-}\sfrac12 mn{-}m{+}4n{+}1\big) 
\qquad\Longrightarrow\qquad
\rho\=\frac{\alpha'}{6\times 24}\,\big(\sfrac{1}{4}m^2{-}\sfrac 12 mn{-}m{+}4n{+}16\big),
\end{equation}
relating the scale $\rho$ to the parameters $m$ and $n$.
Let us note that for $(m,n)=(-4,0)$ this equation gives $\rho=\frac{\alpha'}{6}$, 
while for $(m,n)=(0,0)$ it yields $\rho=\frac{\alpha'}{9}$, 
both in accord with previously obtained results \cite{Lechtenfeld:2010dr}. 
The third of (\ref{EOM_condensate}) implies that 
\begin{equation}\label{eom3}
\alpha'\,\text{tr}\big[|\tilde R|^2-|F|^2\big]\=
\frac{4}{3\rho}\,\big(2n{-}\sfrac12 m{+}1\big)-\frac{12}{r^2}\frac{\alpha'}{r^2} \=
-\frac{1}{3\rho}\,\big(\sfrac14 m^2{-}\sfrac12 mn{+}m{-}4n{+}12\big).
\end{equation}
Note that the latter expression again correctly reproduces 
the results of \cite{Lechtenfeld:2010dr} for the values $(-4,0)$ and $(0,0)$ of the pair $(m,n)$. 
The Bianchi identity (\ref{HBianchi}) may 
also be brought in this form by inserting (\ref{eom4}),
\begin{equation}\label{bianchimn}
\alpha'\,\text{tr}\big[\tilde R\wedge \tilde R-F\wedge F\big]\= 4\,dH \=
\frac{\alpha'}{36\rho}\,\big(\sfrac14 m^2{-}\sfrac12 mn{-}m{+}4n{+}16\big)\,dH.
\end{equation}
Therefore, in order to obtain a solution (either supersymmetric or not) we have to guarantee that 
(\ref{eom4}), (\ref{eom3}) and (\ref{bianchimn}) are satisfied. 
This is necessary but not sufficient. 
Before however considering the remaining equations of motion, let us first try to solve the above ones for appropriate 
choices of $\tilde R$ and $F$.

\paragraph{Scan for solutions with $\mathbf{\tilde\Gamma=\Gamma^-}$.}

Let us begin our investigation by considering that both $\tilde R$ and $F$ are instantons, i.e.\ 
$\tilde R=R^-\big|_\mathfrak m$ and $F=R^-\big|_\mathfrak h$.
Then, comparing (\ref{eom3}) with~(\ref{bianchimn}), Lemma 4.1 of~\cite{Lechtenfeld:2010dr} demands that
\begin{equation}
\sfrac{\alpha'}{4}\,\big(\sfrac14 m^2{-}\sfrac12 mn{-}m{+}4n{+}16\big)\=
3\rho\,\big(\sfrac14 m^2{-}\sfrac12 mn{+}m{-}4n{+}12\big). 
\end{equation}
With (\ref{eom4}) and assuming $\rho>0$, 
this leads to a quadratic relation between $m$ and $n$,
\begin{equation}\label{mnrel0}
m^2-2mn\=4\,(4n-m),
\end{equation}
whose solutions are found to be
\begin{equation}\label{mnrel1}
m\=n-2\pm\sqrt{n^2+12n+4} \qquad\Longleftrightarrow\qquad n\=\frac{m\,(m+4)}{2\,(m+8)}.
\end{equation}
On the other hand, since $\tilde R$ and $F$ are assumed to be instantons, (\ref{BianchiExpressions}) yields
\begin{equation}
\text{tr}\big(\tilde R\wedge \tilde R-F\wedge F\big)\=(2\beta{-}1)\,\rho^{-1} dH.
\end{equation}
This expression may be compared to (\ref{bianchimn}) to give us~\footnote{
In the case $\beta{=}0$, where $\mathfrak h$ is abelian, one may rescale
$F=\lambda R^-\big|_\mathfrak m$ with $\lambda\in\mathbb R$, 
which replaces the left-hand side with $\lambda^2{-}1$.}
\begin{equation}\label{betamn}
2\beta{-}1\=4\sfrac{\rho}{\alpha'}\=\sfrac{1}{18}(4n-m+8)
\end{equation}
with the help of (\ref{eom4}) and (\ref{mnrel0}), 
thus providing a linear relation between $m$ and $n$ which parametrically depends on~$\beta$.
Plotting in the $mn$ plane the two curves corresponding to (\ref{mnrel1}) and~(\ref{betamn}),
it is easy to see that they intersect for $2\beta{-}1\notin\ ]{-}\sfrac43,\sfrac49[\ $.\footnote{
The notation $]a,b[$ to denote open intervals, to distinguish from pairs~$(a,b)$.}
However, since $2\beta{-}1\sim\rho>0$, we are left with the condition
\begin{equation}\label{betabound}
2\beta-1\ \ge\ \sfrac49 \qquad\text{or}\qquad \beta=0 \quad\text{and}\quad \lambda^2\ \ge\ \sfrac{13}{9}.
\end{equation}
Consulting the table in the previous section, this rules out the $\text{SU(2)}^3/\text{SU(2)}$ case
and enforces the choice of $F=\lambda R^-\big|_\mathfrak m$ with a real scaling parameter~$\lambda$ in the 
$\text{SU(3)}/\text{U(1)}{\times}\text{U(1)}$ case, as in~\cite{Lechtenfeld:2010dr}.

\begin{figure}[t]
 \begin{center}
\includegraphics[width=17cm]{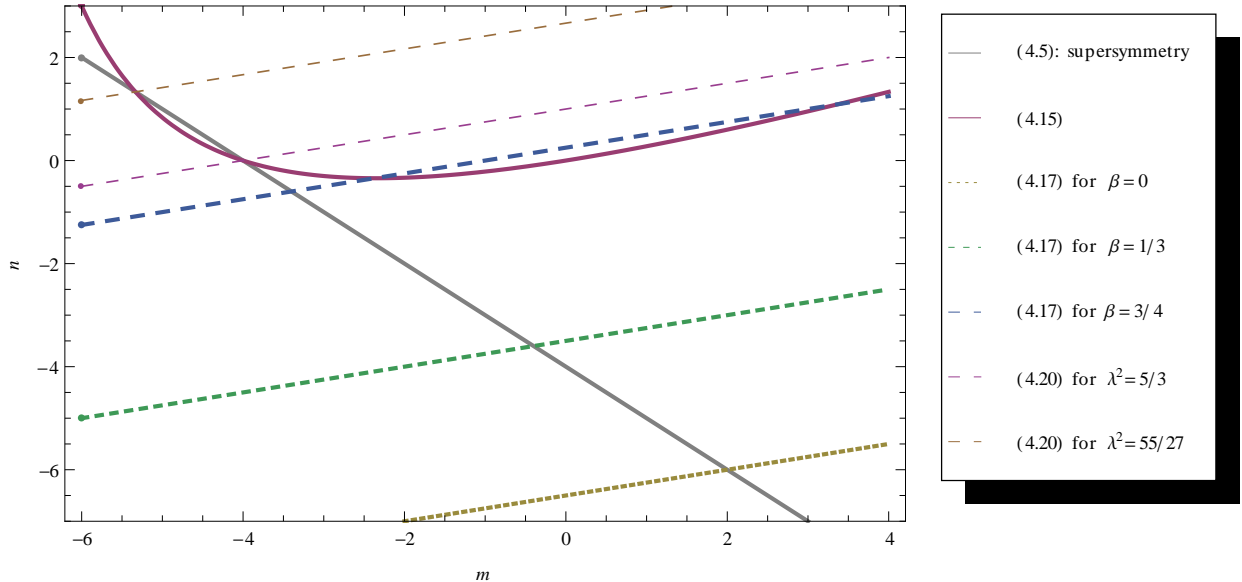}
\caption{Plot in the mn plane of the curves corresponding to equations 
(\ref{c1}),~(\ref{mnrel1}),~(\ref{betamn}) (for special values of $\beta$) and (\ref{btol}) (for special 
values of $\lambda$).}
\end{center}
\end{figure}

Let us examine whether a supersymmetric solution is possible. This means that apart from 
(\ref{mnrel0}) and~(\ref{betamn})
we also have to impose the condition (\ref{c1}). 
The corresponding straight line always intersects with the one from~(\ref{betamn}) and meets the
curve from~(\ref{mnrel0}) in the two points (see Figure~1)
\begin{equation}
(m,n\,|\,2\beta{-}1)=(-4,0\,|\,\sfrac23) \qquad\text{and}\qquad 
(m,n\,|\,2\beta{-}1)=(-\sfrac{16}3,\sfrac43\,|\,\sfrac{28}{27}),
\end{equation}
both for the lower sign in front of the square root in (\ref{mnrel1}). 
Since those $\beta$ values do not appear in table (\ref{betas}), one may try to obtain a solution for
$\text{SU(3)}/\text{U(1)}{\times}\text{U(1)}$ with $F=\lambda R^-\big|_\mathfrak m$.
Replacing $2\beta{-}1$ with $\lambda^2{-}1$, we obtain $\lambda^2=\sfrac53$ and $\lambda^2=\sfrac{55}{27}$,
which are both within the bound (\ref{betabound}) for $\beta=0$. Using (\ref{eom4}), the scales come out as $\rho=\sfrac{\alpha'}{6}$ and $\rho=\sfrac{7\alpha'}{27}$,
respectively. For a supersymmetric solution, these scales should agree with the condition (\ref{c2}), 
which is equivalent to the vanishing of the external gravitino variation. It is readily observed that 
this is indeed the case for the first solution, whereas the second one does not satisfy (\ref{c2}).
The first solution corresponds to $\Sigma=-4H$ and vanishing dilatino and was already obtained 
in~\cite{Lechtenfeld:2010dr}. 
The second one is not a supersymmetric solution.
Therefore we conclude that even with a dilatino condensate it is 
not possible to obtain more than one supersymmetric solution which satisfies all the equations of motion.

Let us now examine the case of non-supersymmetric solutions. 
This means that we do not have to impose (\ref{c1}) or any other condition originating 
from supersymmetry, leaving more freedom in the problem.
We already saw that for any value of $\beta\ge\sfrac{13}{18}$ as well as for $\beta{=}0$ at any value of
$\lambda^2\ge\sfrac{13}{9}$ there is a joint solution to (\ref{mnrel0}) and~(\ref{betamn}), with a
scale $\rho=\sfrac{\alpha'}{4}(2\beta{-}1)\ge\sfrac{\alpha'}{9}$.
Let us perform a case-by-case analysis for the occurring values of $\beta$:
\begin{itemize}
\item 
For $\beta=0$, we must choose $F=\lambda R^-\big|_\mathfrak m$ and obtain
\begin{equation}\label{btol}
\lambda^2-1\=\sfrac{1}{18}(4n-m+8) \quad \Longrightarrow\quad m-4n\=26-18\lambda^2.
\end{equation}
This equation combined with (\ref{mnrel1}) leads to an infinity of possible solutions parametrized by 
$\lambda$. Indeed, we find
\begin{eqnarray}
m\!\!&=&\!\!-13+\ 9\,\lambda^2\pm\sqrt{3}\sqrt{27\lambda^4-30\lambda^2-13}, \\
4n\!\!&=&\!\!-39+27\lambda^2\pm\sqrt{3}\sqrt{27\lambda^4-30\lambda^2-13}.
\end{eqnarray}
A direct observation is that for the borderline value of $\lambda^2=\sfrac{13}{9}$ we obtain $m{=}0$ and $n{=}0$. 
This is exactly the non-supersymmetric solution with vanishing gaugino and dilatino, which was found in 
\cite{Lechtenfeld:2010dr}. Moreover, for the special values of $\lambda^2=\sfrac{5}{3}$ and 
$\lambda^2=\sfrac{55}{27}$ we reproduce the supersymmetric and non-supersymmetric solutions discussed earlier. 
However, here there are obviously more possibilities since any other value of $\lambda^2\ge\sfrac{13}{9}$
yields a non-supersymmetric solution, giving us an infinite family on $\text{SU(3)}/\text{U(1)}{\times}\text{U(1)}$.
\item 
For $\beta=\sfrac34$, we obtain 
\begin{equation}
m-4n=-1,
\end{equation} 
which, combined with (\ref{mnrel1}), leads to the values 
\begin{equation}
(m,n)=\big(\sfrac 12(1\pm\sqrt{33}),\sfrac 18(3\pm\sqrt{33})\big).
\end{equation} 
Then we also find that $\rho=\sfrac{\alpha'}{8}$, which is a legitimate solution.
\item 
For $\beta=\sfrac 13$, we land outside the admissible range, as discussed above. 
\item 
There is a fourth case corresponding to the manifold 
$\text{Sp(2)}/\text{Sp(1)}{\times}\text{U(1)}$, where there is no common 
value for $\beta$. Instead, we have $\beta=0$ on $\mathfrak u(1)$ and $\beta=\sfrac23$ on 
$\mathfrak{sp}(1)$. This allows one to calculate the quantities 
tr$(R^\pm\wedge R^\pm)$, which are no longer proportional to $dH$. 
However, we have the freedom to restrict the curvature $\tilde R$ to the 
$\mathfrak u(1)$ part of~$\mathfrak h$.
Again this can be rescaled, enabling us to satisfy the Bianchi identity
for a particular choice, 
\begin{equation}
\tilde R=R\big|_{\mathfrak u(1)} \und 
F\=R^-\big|_\mathfrak m \qquad\Longrightarrow\qquad
\text{tr}(\tilde R\wedge \tilde R - F\wedge F) \= \sfrac13\,\rho^{-1}dH. 
\end{equation} 
In contrast to the SU(3)/U(1)$\times $U(1) case, 
there are then no free parameters left in the gauge field.
Moreover,
\begin{equation}
\text{tr}(|\tilde R|^2-|F|^2)\= -\sfrac 1{3\rho^2}.
\end{equation}
Then, according to the above, we end up with the condition
\begin{equation}
m-4n\=2,
\end{equation}
which however has no intersection with~(\ref{mnrel0}).
Hence, although we find that $\rho=\sfrac{\alpha'}{12}>0$, there is no solution on 
$\text{Sp(2)}/\text{Sp(1)}{\times}\text{U(1)}$.
\end{itemize}

\paragraph{Scan for solutions with $\mathbf{\tilde\Gamma=\Gamma^+}$.}

Let us continue by choosing $\tilde R=R^+$, which is not an instanton. For $F$ we insist on the previous 
choice, i.e.\ $F$ is still chosen to be an instanton gauge field. In that case (\ref{bianchimn})
and (\ref{eom3}) yield, respectively,
\begin{eqnarray}\label{gammaplus1}
2\beta\!\!&{=}&\!\!4\sfrac{\rho}{\alpha'}\=
\sfrac{1}{36}(\sfrac14 m^2-\sfrac12 mn-m+4n+16),\\[4pt] \label{gammaplus2}
\sfrac 43-2\beta\!\!&{=}&\qquad -\sfrac{\rho}{3\alpha'}(\sfrac14 m^2-\sfrac12 mn+m-4n+12).
\end{eqnarray}
Let us again perform a case-by-case analysis for the occurring values of~$\beta$.
\begin{itemize}
 \item For $\beta=0$, in the present case it is obvious that we are immediately led to $\rho=0$, which is not 
acceptable. However, as we discussed before, one may try to obtain solutions for the rescaled connection 
$F=\lambda R^-|_{\mathfrak m}$. Presently such a choice leads to the replacement of $2\beta$ in 
(\ref{gammaplus1}) and (\ref{gammaplus2}) by $\lambda^2$. Then (\ref{gammaplus1}) requires that 
$\lambda^2\sim\rho>0$, which is true since $\lambda\in\mathbb R$. Solving the two equations we 
obtain the following result,
\begin{eqnarray}
m\!\!&{=}& \, -4 + \ 9\,\lambda^2+ \lambda^{-2}\bigl( \ \; 4 \pm \sqrt{
16 - 96 \lambda^2 + 24 \lambda^4 + 72 \lambda^6 + 81 \lambda^8} \bigr),\\
4n\!\!&{=}&\!\! -12 + 27 \lambda^2 + \lambda^{-2} \bigl( 12 \pm \sqrt{
16 - 96 \lambda^2 + 24 \lambda^4 + 72\lambda^6 + 81 \lambda^8} \bigr),
\end{eqnarray}
under the condition that $|\lambda|$ lies in the domain~\footnote{
The interval border value $0.424$ approximates the positive root of the polynomial equation 
$27\lambda^6+42\lambda^4+36\lambda^2-8=0$; the other border value is $\sfrac13\sqrt{6}\approx0.816$ .}
\begin{equation}
|\lambda|\,\in\, \ ]0\,,0.424[\ \cup\ ]0.8816\,,\infty[\ .
\end{equation}
Therefore we obtain again a infinite family of solutions on $\text{SU(3)}/\text{U(1)}{\times}\text{U(1)}$.

It is interesting to observe that the parameter $n$ does not vanish for any of the allowed values of $\lambda$.
This is in accord with the results of \cite{Lechtenfeld:2010dr}. However, for the value 
$\lambda^2=\sfrac 29(1+\sqrt{10})$, which lies in the allowed domain, the parameter $m$ vanishes. 
Then this solution corresponds to vanishing gaugino but non-vanishing dilatino condensate. 
\item For $\beta=\sfrac34$, the first of the above equations yields $\rho=\sfrac 38\alpha'$, which is 
positive as required. Then we can proceed to the solutions for $m$ and $n$. We find
\begin{eqnarray}
m\!\!&{=}&\!\! \sfrac 16\big(\ \,73\pm\sqrt{9265}\big)\ \approx\ 12.167 \pm 16.042,\\
4n\!\!&{=}&\!\!\sfrac 16\big(219\pm\sqrt{9265}\big) \ \approx\ 36.500 \pm 16.042.
\end{eqnarray}
These values provide two solutions on $G_2/\text{SU(3)}$ for the plus-connection.
\item For $\beta=\sfrac 13$, we obtain $\rho=\sfrac{\alpha'}{6}>0$. Moreover, solving the two equations for 
$m$ and $n$ we arrive at
\begin{equation}
 (m,n)=(8,6),
\end{equation}
which corresponds to a solution on $\text{SU(2)}^3/\text{SU(2)}$ for the plus-connection.
\item As far as the space $\text{Sp(2)}/\text{Sp(1)}{\times}\text{U(1)}$ is concerned, unlike the 
$\Gamma^-$ case where the connection is an $\mathfrak h$-gauge field and therefore the curvature can be restricted 
to the $\mathfrak u(1)$ part of~$\mathfrak h$, this freedom does not exist here. Therefore it is not 
possible to satisfy the Bianchi identity on this space with the plus-connection.
\end{itemize}
Let us note that none of the solutions with plus-connection is supersymmetric since they all fail to satisfy 
the condition~(\ref{c1}).

\paragraph{The remaining equations of motion and fermion masses.} 

Let us finish this section by commenting on the equations of motion which were not treated in detail until 
now. These are the Einstein equation, the gaugino and dilatino equations of motion, the Yang-Mills equation 
and the Kalb-Ramond equation, all appearing in~(\ref{EOM_condensate}).

As far as the (internal) Einstein equation is concerned, all its terms are proportional to $g_{ab}$,\footnote{
For our choice of $\eta$ being an internal Killing spinor, the right-hand side of the third equation
in~(\ref{EOM_condensate}) vanishes identically.}
and therefore it is enough to solve its trace. 
However, the trace is identical to (\ref{eom3}), which was already 
taken into account. Therefore the Einstein equation on $K$ is satisfied. 

The Yang-Mills equation of motion is also satisfied since in the above analysis we assumed that the gauge 
field is always an instanton. As for the Kalb-Ramond equation, since $\Sigma$ and $\Delta$ are both 
proportional to $H$ it is simplified to 
\begin{equation}
 d\ast H=0.
\end{equation}
 Then this is satisfied due to the relations (\ref{NK_defRels}), (\ref{H-OmegaRel}) and (\ref{duality}).

Finally, let us turn to the Dirac equations for the gaugino and the dilatino. We have already considered 
the decomposition of the corresponding spinors in terms of the spinor $\eta$ on the internal space in 
(\ref{chifactor}) and (\ref{lambdafactor}). Moreover, we know that the spinor $\eta$ is covariantly constant 
with respect to the torsionful connection $\nabla^-$, thus 
\begin{equation}
 \mathcal D^-\eta=0 \qquad \Longrightarrow \qquad \big(\mathcal D-\sfrac 18\gamma(H)\big)\,\eta=0,
\end{equation}
and likewise for $\eta^{\ast}$.
The above indicate that it is useful to express the gaugino and dilatino equations 
in terms of ${\mathcal D}^-$. 
We obtain
\begin{eqnarray}
0\!\!&=&\!\!\big(\widehat{\mathcal D}+\mathcal D^-+\sfrac1{24} \gamma(2H+\sfrac 12\Sigma-\sfrac 12\Delta)\big)\,\chi
\= \big(\widehat{\mathcal D}+\sfrac{1}{48}(4{+}m{-}n)\gamma(H)\big)\,\chi, \label{gaugino2}\\[2pt]
0\!\!&=&\!\!\big(\widehat{\mathcal D}+\mathcal D^-+\sfrac1{24} \gamma(2H+\sfrac 18\Sigma) \big)\,\lambda\qquad\quad\!
\= \big(\widehat{\mathcal D}+\sfrac{1}{48}(4{+}\sfrac14m)\gamma(H)\big)\,\lambda. \label{dilatino2}
\end{eqnarray}
After computing
\begin{eqnarray}
\gamma(H)\,
\big(e^{\frac{i\pi}{4}}\widehat\chi{\otimes}\eta+e^{-\frac{i\pi}{4}}\widehat\chi^\ast{\otimes}\eta^\ast\big)
\!\!&=&\!\!-\sfrac{12}{\sqrt{3\rho}}
\big(e^{-\frac{i\pi}{4}}\widehat\chi{\otimes}\eta^\ast+e^{\frac{i\pi}{4}}\widehat\chi^\ast{\otimes}\eta\big), \\[2pt]
\gamma(H)\,
\big(e^{\frac{i\pi}{4}}\widehat\lambda{\otimes}\eta^\ast+e^{-\frac{i\pi}{4}}\widehat\lambda^\ast{\otimes}\eta\big)
\!\!&=&\!\!-\sfrac{12}{\sqrt{3\rho}}
\big(e^{-\frac{i\pi}{4}}\widehat\lambda{\otimes}\eta+e^{\frac{i\pi}{4}}\widehat\lambda^\ast{\otimes}\eta^\ast\big),
\end{eqnarray}
the vanishing of the coefficients of $\eta$ and $\eta^\ast$ in (\ref{gaugino2}) and (\ref{dilatino2}) yields
massive Dirac equations on AdS$_4$,\footnote{
Note that no imaginary unit appears here because of our sign choice in the Clifford algebra (\ref{clifford}).}
\begin{equation}
\widehat{\mathcal D}\,\widehat\chi\=m_\chi\,\widehat\chi^\ast \und
\widehat{\mathcal D}\,\widehat\lambda\=m_\lambda\,\widehat\lambda^\ast,
\end{equation}
with four-dimensional gaugino and dilatino masses given by
\begin{equation}
m_\chi \= \frac{4{+}m{-}n}{4\sqrt{3\rho}} \und
m_\lambda \= \frac{16{+}m}{16\sqrt{3\rho}}.
\end{equation}
With the help of (\ref{eom4}) one may express the scale $\rho$ in terms of $\alpha'$ 
for each case of values~$(m,n)$.
A similar formula was determined in \cite{Manousselis2} for a general choice of connection.
In the following section, where we summarize our results, we shall 
provide the explicit masses in terms of the string slope~$\alpha'$.

\section{Summary of results and discussion}

In this paper we examined AdS${}_4$ heterotic compactifications on nearly K\"ahler manifolds with non-vanishing 
gaugino and dilatino bilinears. We were able to find a set of solutions with properties which are 
summarized in the following three tables. 
In the first table appear the common features of all the solutions we found:
\begin{center}
 \begin{tabular}{l*{4}{c}r}
quantity       & symbol       & value  \\
\hline
AdS radius  & $r$ &  $\sqrt{\sfrac{\alpha'}{2}}$  \\
$H$-flux         &  $H_{abc}$ & $-f_{abc}$   \\
dilaton vev & $\phi$ & constant \\
$10d$ gaugino         & $\chi$ &  
$e^{\frac{i\pi}4}\widehat\chi\otimes\eta + e^{-\frac{i\pi}4}\widehat\chi^*\otimes\eta^*$  \\
$10d$ dilatino   & $\lambda$ &  
$e^{\frac{i\pi}4}\widehat\lambda\otimes\eta^{\ast} + e^{-\frac{i\pi}4}\widehat\lambda^{\ast}\otimes\eta$  \\
$10d$ gravitino  & $\psi$ & 0
\end{tabular} 
\end{center}
In the second table we list some solution-dependent features for the five (sets of) solutions determined:
\begin{center}
 \begin{tabular}{l*{4}{c}r}
 & internal space      & internal scale $\rho$  & metric $g_{ab}$ & connection $\tilde \Gamma$ & gauge field $F_{ab}$ 
 \\
\hline \\
1&$\text{SU(3)}/\text{U(1)}{\times}\text{U(1)}$  & $(\lambda^2-1)\sfrac {\alpha'}4 $ &  
$(1-\lambda^2)\sfrac {3\alpha'}4f^c_{ad}f^{d}_{bc}$ & $\Gamma^-$ & 
$\lambda R_{ab}^-|_{\mathfrak m}$ 
\\ \\
2&$G_2/\text{SU(3)}$         &  $\sfrac{\alpha'}{8}$ & $-\sfrac{3\alpha'}{8}f^c_{ad}f^{d}_{bc}$ & 
$\Gamma^-$ & $R_{ab}^-|_{\mathfrak h}$  \\ \\
3&$\text{SU(3)}/\text{U(1)}{\times}\text{U(1)}$          & $\lambda^2\sfrac{\alpha'}{4}$ &  
$-\lambda^2\sfrac {3\alpha'}4f^c_{ad}f^{d}_{bc}$ 
& $\Gamma^+$ & $\lambda R_{ab}^-|_{\mathfrak m}$\\ \\
4&$G_2/\text{SU(3)}$    & $\sfrac{3\alpha'}{8}$ &  $-\sfrac{9\alpha'}{8}f^c_{ad}f^{d}_{bc}$ & 
$\Gamma^+$ & $R_{ab}^-|_{\mathfrak h}$ \\ \\
5&$\text{SU(2)}^3/\text{SU(2)}$ & $\sfrac{\alpha'}{6}$ & $-\sfrac{\alpha'}{2}f^c_{ad}f^{d}_{bc}$ & 
$\Gamma^+$ & $R_{ab}^-|_{\mathfrak h}$
\end{tabular} 
\end{center}
For the first solution the allowed domain for $\lambda$ is $\lambda^2\ge \sfrac{13}9$, while for the 
third solution it is $\lambda^2\ge \sfrac23$ or $\lambda^2\le 0.18, ~\lambda\ne 0$. 
Finally, the remaining solution properties are given in the third table: 
\begin{center}
 \begin{tabular}{l*{4}{c}r}
 & bilinear values $(m,n)$      & gaugino mass$^2$ $m^2_{\chi}$  & dilatino mass$^2$ $m^2_{\lambda}$ 
 \\
\hline \\
1&  
 $\big(-13+9\lambda^2\pm P(\lambda^2),\sfrac 14(-39+27\lambda^2\pm P(\lambda^2))\big)$ 
& $\sfrac{3 (1 + 3 \lambda^2 \pm P(\lambda^2))^2}{64 \alpha' (\lambda^2-1)}$    & 
 $\sfrac{ (3 + 9\lambda^2 \pm P(\lambda^2))^2}{192 \alpha' (\lambda^2-1)}$ 
\\ \\
2&    $\big(\sfrac 12(1\pm\sqrt{33}),\sfrac 18(3\pm\sqrt{33})\big)$     &  
$\sfrac{33}{64\alpha'}(7 \pm \sqrt{33})$ & $\sfrac{11}{64\alpha'}(17 \pm \sqrt{33})$
   \\ \\
3&   $\big(-4+9\lambda^2+\frac{4 \pm Q(\lambda^2)}{\lambda^2},-3 + \sfrac {27}{4} \lambda^2 
+ \frac{3 \pm \sfrac 14 Q(\lambda^2)}{\lambda^2}\big)$  
& $\sfrac{(4 + 12 \lambda^2 + 9 \lambda^4 \pm 3 Q(\lambda^2))^2}{192 \lambda^6\alpha'}$ &  
$\sfrac{(4 + 12 \lambda^2 + 9 \lambda^4 \pm  Q(\lambda^2))^2}{192 \lambda^6\alpha'}$
\\ \\
4&  $\big(\sfrac 16\big(73\pm\sqrt{9265},\sfrac 1{24}\big(219\pm\sqrt{9265}\big)$  & 
$\sfrac{55973 \pm 507 \sqrt{9265}}{5184\alpha'}$ &  $\sfrac{(169 \pm  \sqrt{9265})^2}{10368 \alpha'}$ 
\\ \\
5& $(8,6)$ & $\sfrac{9}{2\alpha'}$ & $\sfrac{9}{2\alpha'}$ 
\\ \\
\end{tabular} \\ 
with \quad $P(\lambda^2)=\sqrt{3}\sqrt{27\lambda^4-30\lambda^2-13}$ \quad and \quad
$Q(\lambda^2)=\sqrt{ 16 - 96 \lambda^2 + 24 \lambda^4 + 72 \lambda^6 + 81 \lambda^8}$ .
\end{center}

Let us moreover note that all the above solutions are non-supersymmetric, except for the 
first solution on $\text{SU(3)}/\text{U(1)}\times\text{U(1)}$ with the special value $\lambda^2=\sfrac53$ 
and the lower sign, yielding $(m,n)=(-4,0)$, $P(\sfrac53)=6$ and $(m^2_{\chi},m^2_{\lambda})=(0,\sfrac9{8\alpha'})$.

Having summarized our findings let us discuss some interesting aspects of the above solutions. The first 
message is that, for heterotic compactifications on SU(3) structure manifolds, the Killing spinor equations 
and the Bianchi identity {\textit{do not imply}} the equations of motion. 
The only two exceptions are the well-known Calabi-Yau case of vanishing $H$-flux and connection $\Gamma^-$ 
and the first (and only supersymmetric) solution in the above tables on 
$\text{SU(3)}/\text{U(1)}{\times}\text{U(1)}$ again with connection $\Gamma^-$ and $\lambda^2=\sfrac53$. The latter 
corresponds to vanishing dilatino condensate but non-vanishing $H$-flux and gaugino condensate. Turning on 
a non-vanishing dilatino condensate does not improve the situation drastically from the point of view of 
supersymmetric vacua. Indeed, it turns out that it does not lead to any new {\textit{supersymmetric}} solution.
Thus although it provides an extra freedom to solve the Killing spinor equations and the Bianchi identity 
\cite{Manousselis:2005xa}, this freedom is lost at the level of the equations of motion.

It goes without saying that the equations of motion and the absence of anomalies 
are more fundamental than the supersymmetry equations. Therefore, in the present paper 
we worked at the level of the equations of motion and the Bianchi identity at first order in $\alpha'$, 
allowing for non-vanishing supersymmetry variations.~\footnote{
It is worth mentioning that non-supersymmetric Minkowski compactifications with non-vanishing fermion bilinears 
were considered recently in the context of 11d supergravity~\cite{Farakos:2011kk}.} 
Our basic assumptions were threefold:
(1)~the $H$-flux is identified with the torsion of the internal space, 
(2)~the gauge field is a generalized instanton~\footnote{This assumption facilitates the solution 
of the Yang-Mills equations. However, different assumptions for the gauge field are fully legitimate 
as long as the corresponding equations are satisfied. Different choices for the gauge field appear for 
example in \cite{Lukas}, where however the equations of motion are not checked explicitly.} and 
(3)~gaugino and dilatino bilinears may develop a vacuum expectation value. 
It is interesting that under these assumptions it was 
possible to fully classify the consistent vacua of the theory on nearly K\"ahler manifolds, regardless of 
their remnant supersymmetry. This examination resulted in the solutions appearing in the above tables. 
The introduction of a dilatino condensate clearly brings up more solutions than before, albeit 
non-supersymmetric ones.  

A further interesting issue is that all our solutions fix the radius of AdS${}_4$ as well as the volume modulus 
$\rho$ of the internal nearly K\"ahler space. This provides a partial moduli stabilization in four dimensions. 
As far as the rest of the moduli are concerned, i.e.\ the four-dimensional dilaton and the K\"ahler moduli 
of the compactification, their stabilization may be studied within the effective scalar potential in the 
four-dimensional theory. This potential was partially determined for nearly K\"ahler manifolds in 
\cite{Chatzistavrakidis:2008ii,Chatzistavrakidis:2009mh} in the absence of fermion condensates. The 
presence of fermion condensates typically leads to an effective component in the four-dimensional 
superpotential, which is known at least for the gaugino condensates \cite{Dine,Derendinger}. This effective 
superpotential may assist the stabilization of the dilaton in four dimensions \cite{deCarlos}.

Since generically supersymmetry is broken, the gaugino and the dilatino acquire non-vanishing masses. 
These masses depend on the internal geometry and are calculable as we have demonstrated. 
It is worth noting that in the single supersymmetric case on $\text{SU(3)}/\text{U(1)}{\times}\text{U(1)}$ 
the gaugino mass vanishes, as expected, while the dilatino mass squares to $m^2_{\lambda}=\sfrac9{8\alpha'}$. 
As a potentially interesting feature of our analysis, the stabilization of the internal scale allows one 
to calculate the fermion masses explicitly in terms of the string scale.

\paragraph{Acknowledgments.}
This work was partially supported by the Deutsche Forschungsgemeinschaft,
the cluster of excellence QUEST, the Heisenberg-Landau program,
the Russian Foundation for Basic Research, the SFB-Tansregio TR33
``The Dark Universe'' (Deutsche Forschungsgemeinschaft) and
the European Union 7th network program ``Unification in the LHC era'' (PITN-GA-2009-237920).

\end{document}